# A Study on Impact of Environmental Accounting on Profitability of Companies listed in Bombay Stock Exchange

Nandini E.S
*Student, M.Com in Accounting and taxation*
*M.S Ramaiah University of applied Sciences*

Sudharani R
*Assistant Professor*
*M.S Ramaiah University of Applied Sciences*

Dr. Suresh N
*Professor*
*M.S Ramaiah University of Applied Sciences*

**Abstract-** The study focuses on the Impact of Environmental Accounting on Profitability of Companies listed in Bombay Stock Exchange. The study has considered the Amount spent on Environmental protection as Independent variable and Return on Capital Employed, Return on Assets, Return on Net worth/equity, Net Profit Margin and Dividend per Share as Dependent variable. The present study is to analyses the relationship between Amounts spent on Environmental protection cost and Return on Capital Employed, Return on Assets, Return on Net worth/equity, Net Profit Margin and Dividend per Share. The data is collected of 18 companies listed in Bombay Stock Exchange of 10 years from Annual report of companies. The data collected were analysed using Panel data Regression in E-Views. Results revealed that there is a significant Relationship between Environmental protection Cost and Return on Capital Employed, Return on Assets, and Return on Net worth/equity, Net Profit Margin and Dividend per Share. The study shows that Environmental accounting impact positively on Firms profitability.

## 1. INTRODUCTION

Environmental accounting is a broader term that relates to the provision of environmental-performance-related information to stakeholders both within, and outside, an organization. Environmental accounting is defined in these guidelines, aims at achieving sustainable development, maintaining a favorable relationship with the community, and pursuing effective and efficient environmental conservation activities. These accounting procedures allow a company to identify the cost of environmental conservation during the normal course of business, identify benefit gained from such activities. Environmental accounting can more accurately identify true costs by clarifying the environmental impacts caused by material acquisition and processing, manufacturing, sales, distribution, use, maintenance, and disposal. It can help companies and organizations develop innovative solutions to change resource use and eliminate resource constraints, meet regulatory requirements, and avoid ecological crises. It can also provide consumers with the additional information they need to make more informed purchasing choices. There are several advantages environmental accounting brings to business; notably, the complete costs, including environmental remediation and long term environmental consequences and externalities can be quantified and addressed.

### 1.1 Necessity of Environmental Accounting

Accounting for environment has become increasingly relevant to enterprises because issue of the availability of natural recourses and pollution of the environment has become the subject of economic, social and political debate throughout the world. Steps are being taken at the national and international level to protect the environment and to reduce, prevent and mitigate the effect of pollution. As a result there is a trend for the enterprise to disclose the community large data related to environment policies, environment management programmes and the impact of environment performance on their financial performance. Carrying out environmental conservation activities, a company or other organizations can accurately identify and measure investments and costs related to environmental conservation activities, and can prepare and analyse this data. By having better insight into the potential benefit of these investments and costs, the company can not only improve the efficiency of its activities, but environmental accounting also plays a very important role in supporting rational decision-making. Companies and other organizations are required to have accountability to stakeholders, such as consumers, business partners, investors and employees, when utilizing environmental resources, i.e. public goods, for their business activities. Disclosure of environmental accounting information is a key process in performing accountability. Consequently, environmental accounting helps companies and other organizations boost their public trust and confidence and are associated with receiving a fair assessment.

## 2. LITERATURE REVIEW

Daniel Mogaka Makori Ambrose Jagongo, PhD (2013) This study is to establish whether there is any significant relationship between environmental accounting and profitability of selected firms listed in India. The data for the study were collected from annual reports and accounts of 14 randomly selected quoted companies in Bombay Stock Exchange in India. The data were analysed using multiple regression models. The key findings of the study shows that there is significant negative relationship between Environmental Accounting and Return on Capital Employed (ROCE) and Earnings per Share (EPS) and a significant positive relationship between Environmental Accounting and Net Profit Margin and Dividend per Share. Based on this it was recommended that government should give tax credit to organizations that comply with its environmental laws and that environmental reporting should be made compulsory in India so as to improve the performance of organizations and the nation as a whole.

Dr. Rabindra Kumar Swain, Roji Kanungo, Sakti Ranjan Dash (2017) This paper attempts to study the conceptual aspects of corporate environmental reporting and its regulatory framework. It also examines the consistency of Indian corporate in disclosing environmental factors as per GRI guidelines, as GRI is the  popularly adopted guideline among the corporate world. It also makes an attempt to identify the extent of environmental disclosure under GRI guidelines by sample companies. For the purpose of the study top 50 Indian companies of on the basis of market capitalisation which are listed in the Bombay stock exchange, have been taken as sample and their annual report for the year 2014-15 have been analyzed to reveal the consistency of disclosure of environmental aspects and to find out the extent to which they are disclosing these aspects. Statistical tools like coefficient of variation, proportion test and chi square test has been used for analyzing the objectives.

Charles Emenike Ezeagba ,John-Akamelu Chitom Rachael ,Umeoduagu Chiamaka(2017)The study examined the relationship between environmental accounting disclosures and return on equity of food and beverage companies in Nigeria. . Data for the study were collected through secondary sources and analyzed using Pearson's correlation statistical technique and multiple regression, with the aid of SPSS version 20.00. The study revealed that there is a significant relationship between environmental accounting disclosures and ROE. It also revealed (-) relationship between environmental accounting disclosures ROCE and NPM

Dr.C.Sengottuvel (2018) Environment accounting involves the identification, measurement and allocation of environmental costs, and the integration of these costs into business and encompasses the way of communicating such information to companies' stakeholders. In this sense, it is a comprehensive approach to ensure good corporate governance that includes transparency in its societal activities. The unserious attitudes of several forms not take environmental accounting into consideration makes performance below expectation. This is because environmental accounting helps the form to record all environmental costs incurred by the business thereby finding a way of reducing the cost (environmental expenses) so that the business can increase profit. The company's performance has been evaluated by analyzing its financial capability. A study on the environmental accounting and firm's profitability analysis of Bannariamman Spinning Mills has been found to be apt in this context which will throw light on the causes of fluctuations in performance.

Amaechi Patrick Egbunike and Godsday Edesiri Okoro (2018) This paper seeks to investigate whether environmental accounting matters to the profitability of Nigerian firms or not. Towards achieving this, an expo-facto research design was adopted and ten non-consumer goods firms listed on the Nigerian Stock Exchange were selected during 2012-2016. The data were sourced from the annual reports and accounts of the selected non-consumer goods firms. The study revealed that there was no significant relationship between environmental accounting and profitability measures among the non-consumer goods firms.

Priyanka Aggarwal (2013) The purpose of this study is to analyze the relationship between environmental responsibility and financial performance of firm through review of extant literature, so as to find answer to the research question 'whether going green is profitable for firm or not'. The results are mixed, inconsistent and often contradictory; ranging from positive, to negative, to statistically insignificant relationship; depending upon the choice of measure of environmental responsibility, measure of financial performance, sample composition, time-period and control variables. This paper attempts to critically analyze prior studies in order to build up scope for further research, so that future researchers may reach to better and more consistent results.

### 3. PROBLEM FORMULATION

**Identification of Research Gap**

After going through several literatures related to the study, it was shown that most of the studies were limited to short durations limited to two to five years with very minimal number of variables. Most of the researches are being conducted on studying the relationship of Environmental Accounting and Profitability of companies using excel spread sheet , SPSS and other statistical tools and none of the researcher have used Panel data regression for studying the same.

Considering the above stated research gaps, the current study have improved and is being conducted using different statistical tools which are not used before to study and analyse the relationship between Environmental Accounting and Profitability for the period of ten years. The current study also focus on analysing large number of companies data compared to earlier studies.

### 4. OBJECTIVES

a) To study and explore various Environmental Accounting factors impacting on Profitability of Companies.
b)  To analyse the Relationship between Environmental Accounting and Return on Capital Employed and Return on Assets.
c) To analyse the Relationship between Environmental Accounting and Return on Net worth/Equity, Net Profit Margin and Dividend per Share.
d)  To provide suitable suggestions based on study finding.

### 5. PROBLEM SOLVING

**5.1 Data Collection**

This study is on the secondary data completely. The required data sets were the annual environmental protection cost, Return on capital employed, Net profit margin, Return on Net worth/equity, Return on assets and dividend per share of the 18 BSE listed companies for the period between 2009 to 2018.
For, first objective paper, articles and journals are been studied available in various websites like Google Scholar.
This study has been done using publicly available secondary data from Annual reports of the selected company.
In second objective and third objective Environmental protection cost is collected from Annual reports and other dependent variables amount has been collected from financial reports of the companies using the reports available in Money control Website.

### 5.2 HYPOTHESIS

H0- There is a significant relationship between Environmental Accounting and Return on Capital Employed, Return on Assets, Return on networth/equity, Net profit margin and Dividend per share.

H1-There is no significant relationship between Environmental Accounting and Return on Capital Employed, Return on Assets, Return on networth/equity, Net profit margin and Dividend per share

### 5.3 Results and discussion

**Results of fixed effect model:**

```
Dependent Variable: ROCE
Method: Panel Least Squares
Date: 03/11/20   Time: 11:27
Sample: 2009 2018
Periods included: 10
Cross-sections included: 18
Total panel (unbalanced) observations: 174
```

| Variable | Coefficient | Std. Error | t-Statistic | Prob. |
|---|---|---|---|---|
| C | 23.92015 | 0.902366 | 26.50826 | 0.0000 |
| EPC | -0.018934 | 0.023307 | -0.812379 | 0.4178 |

Effects Specification

Cross-section fixed (dummy variables)

| | | | |
|---|---|---|---|
| R-squared | 0.890833 | Mean dependent var | 23.45791 |
| Adjusted R-squared | 0.878156 | S.D. dependent var | 26.46666 |
| S.E. of regression | 9.238508 | Akaike info criterion | 7.387398 |
| Sum squared resid | 13229.25 | Schwarz criterion | 7.732353 |
| Log likelihood | -623.7037 | Hannan-Quinn criter. | 7.527333 |
| F-statistic | 70.26912 | Durbin-Watson stat | 0.601656 |
| Prob(F-statistic) | 0.000000 | | |

**Results of random effect model:**

```
Dependent Variable: ROCE
Method: Panel EGLS (Cross-section random effects)
Date: 03/12/20  Time: 12:57
Sample: 2009 2018
Periods included: 10
Cross-sections included: 18
Total panel (unbalanced) observations: 174
Swamy and Arora estimator of component variances
```

| Variable | Coefficient | Std. Error | t-Statistic | Prob. |
|---|---|---|---|---|
| C | 24.16662 | 6.215483 | 3.888132 | 0.0001 |
| EPC | -0.020089 | 0.023093 | -0.869899 | 0.3856 |

Effects Specification

| | S.D. | Rho |
|---|---|---|
| Cross-section random | 26.09286 | 0.8886 |
| Idiosyncratic random | 9.238508 | 0.1114 |

Weighted Statistics

| | | | |
|---|---|---|---|
| R-squared | 0.004416 | Mean dependent var | 2.661092 |
| Adjusted R-squared | -0.001373 | S.D. dependent var | 9.200318 |
| S.E. of regression | 9.201299 | Sum squared resid | 14562.19 |
| F-statistic | 0.762857 | Durbin-Watson stat | 0.546866 |
| Prob(F-statistic) | 0.383654 | | |

Unweighted Statistics

| | | | |
|---|---|---|---|
| R-squared | 0.005864 | Mean dependent var | 23.45791 |
| Sum squared resid | 120473.1 | Durbin-Watson stat | 0.066102 |

**Results of Hausman test:**

```
Correlated Random Effects - Hausman Test
Equation: Untitled
Test cross-section random effects

Test Summary                    Chi-Sq. Statistic    Chi-Sq. d.f.    Prob.
Cross-section random                   0.134906              1         0.7134

Cross-section random effects test comparisons:

Variable           Fixed          Random        Var(Diff.)       Prob.
  EPC           -0.018934       -0.020089        0.000010        0.7134

Cross-section random effects test equation:
Dependent Variable: ROCE
Method: Panel Least Squares
Date: 03/12/20   Time: 12:58
Sample: 2009 2018
Periods included: 10
Cross-sections included: 18
Total panel (unbalanced) observations: 174

Variable         Coefficient     Std. Error      t-Statistic      Prob.
   C              23.92015        0.902366        26.50826        0.0000
  EPC            -0.018934        0.023307       -0.812379        0.4178

                          Effects Specification
Cross-section fixed (dummy variables)

R-squared              0.890833    Mean dependent var      23.45791
Adjusted R-squared     0.878156    S.D. dependent var      26.46666
S.E. of regression     9.238508    Akaike info criterion    7.387398
Sum squared resid      13229.25    Schwarz criterion        7.732353
Log likelihood        -623.7037    Hannan-Quinn criter.     7.527333
F-statistic            70.26912    Durbin-Watson stat       0.601656
Prob(F-statistic)      0.000000
```

**Hypothesis to accept the model:**

**Null hypothesis ($H_0$)**- Random effects model is appropriate

**Alternative hypothesis ($H_a$)**- Fixed effects model is appropriate

The probability value here is 0.7134 which is more than 5% thereby we accept null hypothesis and conclude that the random effect model is appropriate.
This results shows that random effect model is appropriate model to test the relationship between environmental protection cost and return on capital employed.

According to random effect model the probability value is 0.001 which is less than 5%. So here by we reject $H_0$ and accept $H_1$, which means there is a strong relationship between environmental protection cost and return on capital employed.

6.              **CONCLUSION**

The main aim of the study is to know the impact of environmental accounting on firm's profitability. Since the disclosures of environmental information are voluntary, there is a diversity of reporting practice. Large companies tend to report more environment information in their annual reports than the medium-scale businesses; and the disclosure, tend to be more qualitative than quantitative despite the fact that there is a significant relationship between environmental accounting and Firm Profitability. The study shows the impact of environmental accounting on Firms profitability and there is a significant relationship between environmental accounting and Return on capital employed, return on net worth, net profit margin, return on assets and dividend per share. Corporate organizations on their part should ensure that they comply with the environmental laws of the nation as it will go a long way in enhancing their performances

**7. LIMITATIONS OF THE STUDY**

- The main Research Gap for the study is to know the impact of Environmental Accounting and Profitability of Companies listed in Bombay Stock Exchange.
- The study was limited for ten years with one independent variable and five dependent variable
- The study was conducted for 18 companies listed in BSE, further study can be conducted for more number of companies concentrating on different type of industries.

REFERENCES

[1] Makori, D.M. and Jagongo, A., 2013. Environmental accounting and firm profitability: An empirical analysis of selected firms listed in bombay stock exchange, India. International Journal of Humanities and Social Science, 3(18), pp.248-256.

[2] Tapang, A.T., Bassey, E.B. and Bessong, P.K., 2012. Environmental activities and its implications on the profitability of oil companies in Nigeria. International Journal of Physical and Social Sciences, 2(3), pp.285-302.

[3] Paper Swain, R.K., Kanungo, R. and Dash, S.R., Environmental Disclosure Practices in India: Evidence from Top 50 Companies of Bse. International Organization for ScientificResearch, 9(9), pp.05-14.

[4] Priyanka Aggarwal(2013)Relationship between Environmental Responsibility and Financial Performance of Firm: A Literature Review IOSR Journal of Business and Management PP 13-22 Volume 13

[5] Dr.C.Sengottuvel Assistant (2018) ENVIRONMENTAL ACCOUNTING AND FIRMS PROFITABILITY International Journal of Innovative Research in Management Studies. pp.22-27. Volume 3,

[6] Egbunike, Amaecgi & Okoro, Godsday.(2018).Does green accounting matter to the profitability of firms? A canonical assessment. Ekonomski horizonti. 20. 17-26. 10.5937

[7] Ezeagba, C.E., Rachael, J.A.C. and Chiamaka, U., 2017. Environmental accounting disclosures and financial performance: a study of selected food and beverage companies in Nigeria (2006-2015). International Journal of Academic Research in Business and Social Sciences, 7(9), pp.162-174.

[8] Schaltegger, Stefan, and Roger Burritt. Contemporary environmental accounting: issues, concepts and practice. Routledge, 2017.

[9] Ahnad, Yusuf J., and Ernst Lutz. "Environmental accounting for sustainable development." The World Bank Symposium/The World Bank.–1989.–118 p. 1989.

[10] De Beer, Patrick, and Francois Friend. "Environmental accounting: A management tool for enhancing corporate environmental and economic performance." Ecological Economics 58.3 (2006): 548-560.

[11] Yakhou, Mehenna, and Vernon P. Dorweiler. "Environmental accounting: an essential component of business strategy." Business Strategy and the Environment 13.2 (2004): 65-77.

[12] Rubenstein, Daniel Blake. Environmental accounting for the sustainable corporation: strategies and techniques. Westport, CT: Quorum Books, 1994.

[13] Wahyuni, Dina. "Environmental management accounting: Techniques and benefits." Jurnal Akuntansi Universitas Jember 7.1 (2009): 23-35.

[14] United States. Environmental Protection Agency. Office of Pollution Prevention, and ICF Incorporated. An introduction to environmental accounting as a business management tool: key concepts and terms. US Environmental Protection Agency, Office of Pollution Prevention and Toxics, 1995.

[15] Boyd, James, and Spencer Banzhaf. "What are ecosystem services? The need for standardized environmental accounting units." Ecological economics 63.2-3 (2007): 616-626.

[16] Morgenstern, Richard D., William A. Pizer, and Jhih-Shyang Shih. "The cost of environmental protection." Review of Economics and Statistics 83.4 (2001): 732-738.

[17] Lampe, Marc, Seth R. Ellis, and Cherie K. Drummond. "What companies are doing to meet environmental protection responsibilities: Balancing legal, ethical, and profit concerns." Proceedings of the International Association for Business and Society. Vol. 2. 1991.

[18] Sueyoshi, Toshiyuki, and Mika Goto. "Can environmental investment and expenditure enhance financial performance of US electric utility firms under the clean air act amendment of 1990?." Energy Policy 37.11 (2009): 4819-4826.